\documentclass[twocolumn, showpacs,nofootinbib]{revtex4}
\usepackage{epsfig}
\usepackage{amssymb}
\usepackage{amsmath}
\usepackage{amsbsy}
\usepackage{color}
\usepackage{ulem}
\usepackage{epstopdf}

\begin{document}
\title{Dynamical Scaling of Polymerized Membranes}

\author{Ken-ichi Mizuochi, Hiizu Nakanishi and Takahiro Sakaue}
\email{sakaue@phys.kyushu-u.ac.jp}

\affiliation{Department of Physics, Kyushu University 33, Fukuoka 812-8581, Japan}

\begin{abstract}
Monte Carlo simulations have been performed to analyze the sub-diffusion dynamics of a tagged monomer in self-avoiding polymerized membranes in the flat phase. 
By decomposing the mean square displacement into the out-of-plane ($\parallel$) and the in-plane ($\perp$) components, we obtain good data collapse with two distinctive diffusion exponents $2 \alpha_{\parallel} = 0.36 \pm 0.01$ and $2 \alpha_{\perp} = 0.21 \pm 0.01$, and the roughness exponents $\zeta_{\parallel} = 0.6 \pm 0.05$ and $\zeta_{\perp} = 0.25 \pm 0.05 $, respectively for each component. Their values are consistent with the relation from the rotational symmetry.
We derive the generalized Langevin equations to describe the sub-diffusional behaviors of a tagged monomer in the intermediate time regime where the collective effect of internal modes in the membrane dominate the dynamics to produce negative memory kernels with a power-law. We also briefly discuss how the long-range hydrodynamic interactions alter the exponents.
\end{abstract}

\maketitle

\section{Introduction}
The polymerized membrane belongs to a class of membranes that consist of a two-dimensional network of monomers with fixed connectivity~\cite{NPW}. There are several sheet-like macromolecules in nature modeled by the polymerized membrane including, for example, exfoliated sheets of graphite oxide~\cite{graphite1,graphite2}, clay platelets dispersed from its layers~\cite{clay}, polymerized amphiphillic bilayers/monolayers~\cite{polymerized_layer1,polymerized_layer2}  and the spectrin network of red blood cells~\cite{spectrin}. What makes them distinct from fluid membranes is the permanent connectivity of monomers~\cite{KKN, KKN_PRA}. In this sense, the polymerized membrane can be thought as a generalization of linear polymers with the internal dimension ${\mathcal D}=1$ to the higher degree of connectivity with ${\mathcal D}=2$. It is now well recognized that a number of its unique properties arises from an in-plane elastic degrees of freedom due to the fixed connectivity, and its coupling with the out-of-plane undulation mode~\cite{Nelson1990}. Among others, the most notable is the existence of a flat phase, in which the surface normals exhibit a long range order, a feature never expected for linear polymers nor fluid membranes, at finite temperature. Surprisingly enough, once the self-avoiding constraints between monomers set in,  such a flat phase realizes even for flexible membranes, which have no explicit bending rigidity in their microscopic Hamiltonian~\cite{Nelson1990}.

Turning our attention to the dynamics, there have been relatively a few attempts so far to analyze it from the standpoint of polymer dynamics~\cite{KKN_PRA, Muth, Frey, Panja_PM, Pandey, Milchev, Stark}. Very recently, the dynamical internal mode of the phantom polymerized membrane has been analyzed, from which the mean square displacement (MSD) of monomers can be deduced~\cite{Panja_PM}. For more realistic membranes with the self-avoidance, numerical simulations have observed sub-diffusive scalings for the monomer MSD in the intermediate time scale~\cite{Pandey, Milchev, Stark}. These works emphasize the complexity in the membrane dynamics compared to the polymer~\cite{Pandey, Milchev}. 

In this paper, we shall show that the membrane dynamics is also amenable to a simple argument developed for the polymer dynamics, provided that the anisotropy in the flat phase is properly taken into account. In particular, we construct the dynamical scaling scenario for the MSD of a monomer in the polymerized membrane.  Using Monte Carlo (MC) simulation, we provide the first evidence that the monomer MSD in the in-plane and the out-of-plane directions follow distinctive scaling behaviors, and demonstrate an excellent agreement with the dynamical scaling predictions. We then propose a generalized Langevin equation to describe the motion of a tagged monomer in the membrane; The equation elucidates the underlying physics behind the observed sub-diffusion behavior through the memory effect  created due to the connectivity in the system~\cite{Panja_GLE, Sakaue}.

\begin{figure}[h]
 \begin{center}
\includegraphics[width=0.45\textwidth]{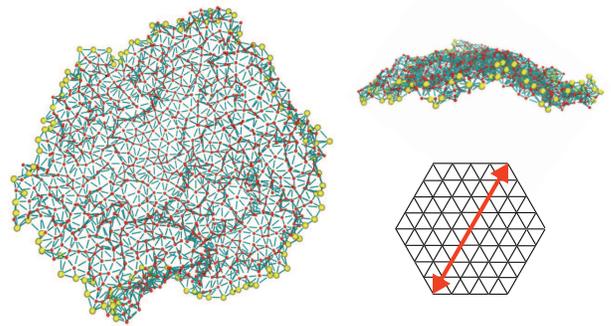}
\caption{Typical snapshot (top and side views) of the membrane with the size $M=38$. Red and yellow balls represent monomers inside the network and those at the periphery, respectively. The linear size $M$ is defined to be the number of bond along the diagonal as shown in the right bottom, where a hexagonal network structure with $M=8$ is illustrated.}
\label{Fig1}
\end{center}
\end{figure}

\section{Monte Carlo Simulation}
We adopt a standard model of a tethered surface~\cite{Kantor_Kremer}, where a hexagonal sheet with a linear diagonal size $M$ is made from monomers in a  $2 {\mathcal D}$ triangular lattice (Fig.~\ref{Fig1}). The total number of monomers is $N = (3M^2+6M + 4)/4$ (note that $M$ is even). The tethering is enforced by a simple rigid potential between nearest-neighbor monomers
\begin{eqnarray}
V_{NN} (r)= \left\{
           \begin{array}{ll}
              0 &  \mbox{ (for $r < b$) }\\
              \infty &  \mbox{ (for $ r > b$) }
           \end{array}
        \right. \label{V_NN}
\end{eqnarray}
and the self-avoidance is imposed by a hard-sphere potential which acts on all monomer pairs
\begin{eqnarray}
V_{SA} (r)= \left\{
           \begin{array}{ll}
              \infty &  \mbox{ (for $r < \sigma$) }\\
              0 &  \mbox{ (for $ r > \sigma$) }
           \end{array}
        \right. \label{V_SA}
\end{eqnarray}
where we set $\sigma = (4/9) b$.
In each MC trial, we attempt to update the position $r_{\alpha}$ ($\alpha=1 \sim 3$) of a randomly selected monomer according to $r_{\alpha} \rightarrow r_{\alpha} + \Delta r_{\alpha}$, where $\Delta r_{\alpha}$ is a random displacement with the magnitude $|\Delta r_{\alpha}|=(1/18)b$. The move is accepted only if the new position is allowed by the constraints given by Eqs.~(\ref{V_NN}) and~(\ref{V_SA}). One MC step consists of a sequence of $N$ such MC trials. All the measurements are made after the equilibration of the membrane conformation by relaxing the system longer than its relaxation time.

To determine the shape and the orientation of membrane, we calculate the shape tensor
\begin{eqnarray}
S_{\alpha \beta} = \frac{1}{N}\sum_{i=1}^{N}(r_{i, \alpha}-r_{\alpha}^{(CM)})(r_{i, \beta}-r_{\beta}^{(CM)}), 
\end{eqnarray}
 where $r_{\alpha}^{(CM)}$ is the $\alpha$-component of the center-of-mass position vector.
 
 \begin{figure}[h]
 \begin{center}
\includegraphics[width=0.3\textwidth]{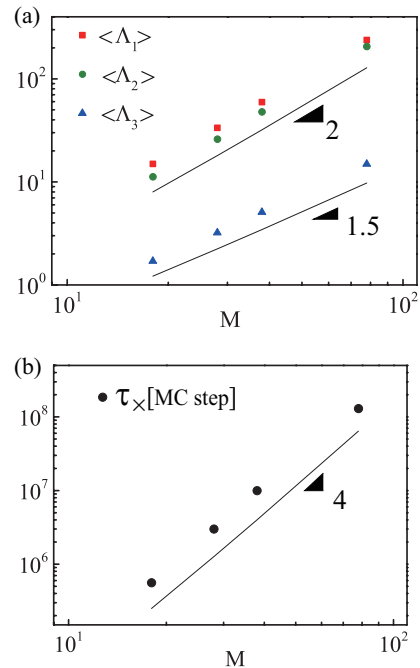}
\caption{System size dependence of  (a) eigenvalues $\langle \Lambda_i \rangle$ $(i=1,2,3)$ of shape tensor and (b) correlation time $\tau_{\times}$ of the rotational motion.}
\label{Fig2}
\end{center}
\end{figure} 
 
Figure~\ref{Fig2} (a) demonstrate the flatness of the membrane by observing that the largest two eigenvalues $\Lambda_1$ and $\Lambda_2$ of $S_{\alpha \beta}$ are similar in magnitude with the same system size dependence $\langle \Lambda_1 \rangle \sim \langle \Lambda_2 \rangle \sim M^{2\nu}$ with $\nu \simeq 1$, while the smallest eigenvalue $\Lambda_3$ is much smaller with the distinct scaling $\langle \Lambda_3 \rangle \sim M^{2 \zeta_{\parallel}'}$ with an apparent roughness exponent $\zeta_{\parallel}' \simeq 0.75$, where $\langle \cdots \rangle$ indicates the ensemble averaging~\footnote{\label{fn1}The effective roughness exponent $\zeta_{\parallel}'$ obtained in this way is affected by the edge effect in open membranes~\cite{Boal, Milchev,Nelson1990,Gompper}.  The elimination of the edge effect leads to the smaller value $\zeta \simeq 0.6$~\cite{Zhang, Petsche_Grest}, which is in accordance with our dynamical analysis below.}. 
Let ${\vec n}(t)$ denote the normalized eigenvector corresponding to $\Lambda_3$, and adopt it as an operative definition of the instantaneous membrane normal at time $t$.
Figure~\ref{Fig2} (b) shows the system size dependence of  the correlation time $\tau_{\times}$ of the time correlation function of the normal vector, which decays almost exponentially
\begin{eqnarray}
\langle {\vec n}(t+t_0) \cdot {\vec n}(t_0) \rangle \simeq e^{-t/\tau_{\times}}, 
\label{tau_nn}
\end{eqnarray}
The logarithmic plot clearly shows that $\tau_{\times}$ scales with the system size as
\begin{eqnarray}
\tau_{\times} \sim M^{4},
\end{eqnarray}
which means that large membranes maintain their orientation for rather long time.
Note that the above scaling for $\tau_{\times}$ accords with the rotational diffusion time of a disk-like molecule under the free-draining dynamics.

We now tag the monomer at the center of the network, and track its trajectory ${\vec r}^*(t)$ for a long time.
We decompose the displacement of the tagged monomer $ {\vec R}(t) \equiv {\vec r}^*(t+t_0)-{\vec r}^*(t_0)$ during the time interval $t$ into the out-of-plane component ${\vec R}_{\parallel}(t) \equiv ({\vec n}(t_0) \cdot {\vec R}(t)){\vec n}(t_0)$ and the in-plane component ${\vec R}_{\perp}(t) \equiv {\vec R}(t)- {\vec R}_{\parallel}(t)$~\footnote{Note that we adopt the convention that the subscripts $\parallel$ and $\perp$ represent, respectively, the direction parallel and perpendicular to the membrane normal ${\vec n}$.}. 
This leads to the decomposition of MSD as $\langle ({\vec R} (t)) ^2 \rangle = \langle ({\vec R}_{\parallel} (t)) ^2 \rangle + \langle ({\vec R}_{\perp} (t)) ^2 \rangle$.

\begin{figure}[h]
 \begin{center}
\includegraphics[width=0.4\textwidth]{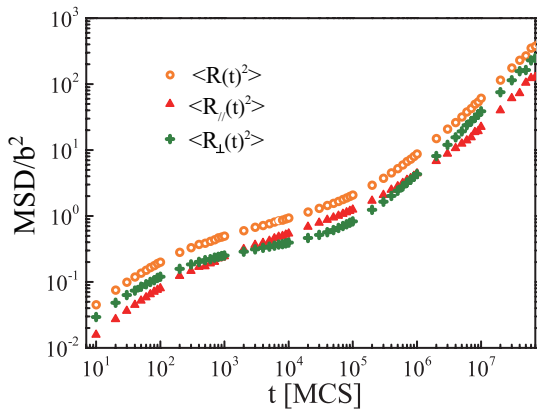}
\caption{MSD of the tagged monomer of the polymerized membrane ($M= 38$) as a function of time (in unit of MC step), and its decomposition into the out-of-plane component $R_{\parallel}$ and the in-plane component $R_{\perp}$.}
\label{Fig3}
\end{center}
\end{figure} 

In Fig.~\ref{Fig3}, MSD of the tagged monomer together with its out-of-plane and in-plane components are plotted as a function of time for the membrane size $M=38$.
In a very short time scale $( t \ll 10^2)$, the monomer exhibits almost normal diffusion, since the motion is yet hardly affected by the connectivity in the network. At $t \gtrsim 10^2$, however, the motion starts to slow down substantially, marked by the onset of distinct dynamics between the out-of-plane and the in-plane directions.

\begin{figure}[h]
 \begin{center}
\includegraphics[width=0.4\textwidth]{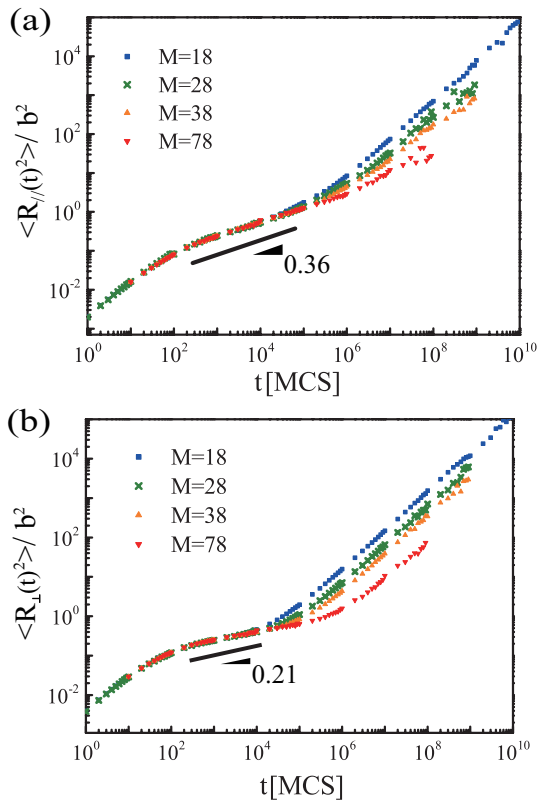}
\caption{MSD of the tagged monomer of the polymerized membranes of various sizes ($M= 18,\ 28,\ 38$ and $78$) as a function of time (in unit of MC step). (a) the out-of-plane component $R_{\parallel}$, and (b) the in-plane component $R_{\perp}$. The numbers in the graphs are the slopes $2 \alpha_{\parallel}$ and $2 \alpha_{\perp}$ in the intermediate time regime.}
\label{Fig4}
\end{center}
\end{figure} 

In Fig.~\ref{Fig4}, we plot the out-of-plane and the in-plane components of the MSD of the tagged monomer in the membranes of various sizes. It is clearly seen that the behaviors in the short and the intermediate time scales are independent of the system size. The intermediate scale dynamics can be characterized by the out-of-plane and the in-plane anomalous diffusion exponents $2\alpha_{\parallel} = 0.36 \pm 0.01$ and $2\alpha_{\perp} = 0.21 \pm 0.01$, with the definition of the exponents $\langle ({\vec R}_{\parallel} (t)) ^2 \rangle \sim t^{2\alpha_{\parallel}}$ and $\langle ({\vec R}_{\perp} (t)) ^2 \rangle \sim t^{2\alpha_{\perp}}$.
Such sub-diffusive behaviors persist up to the terminal times $\tau_{\parallel}$ and $\tau_{\perp}$, after which the diffusion again becomes normal.

As seen in Fig.~\ref{Fig3}, in the normal diffusion regimes in both the short and the long time scales, the MSD in the in-plane component is about twice as large as that in the out-of-plane component, reflecting the number of freedoms in respective components. This ordering reverses, however, in the sub-diffusive regime, where the constraint imposed by the connectivity shows up most significantly.  This observation indicates $\tau_{\parallel} > \tau_{\perp}$.


\section{Dynamical Scaling}
The relevant length and time scales in the system would be deduced from the hydrodynamic description of self-avoiding polymerized membranes in their flat phase~\cite{NPW,Nelson1990}.
The fluctuations of the membrane in flat phase can be decomposed into the out-of-plane undulation mode $u_{\parallel}({\vec x}, t)$ and the in-plane phone mode ${\vec u}_{\perp}({\vec x}, t)=(u_1({\vec x}, t), u_2({\vec x}, t))$, where ${\vec x}=(x_1, x_2)$ labels the internal position of the monomer in the ${\mathcal D}=2$ connectivity space.
The long wavelength static properties of the flat phase can be described by the following effective free energy functional~\cite{Nelson1990}
\begin{eqnarray}
{\mathcal H}&=& \frac{1}{2}\int \frac{d^2 q}{(2 \pi )^2}  \Big[ \kappa(q) q^4|u_{\parallel}({\vec q})|^2 \nonumber \\
&&+ \mu(q) q^2|u_{\perp}(q)|^2 + \left\{\mu(q) + \lambda(q) \right\}|{\vec q}\cdot{\vec u}_{\perp}({\vec q})|^2 \Big],  
\end{eqnarray}
in terms of Fourier modes $u_{\parallel}({\vec q}) = \int_{-\infty}^{\infty} d^2x \ e^{i {\vec q} \cdot {\vec x}} u_{\parallel}({\vec x})$ and ${\vec u}_{\perp}({\vec q}) = \int_{-\infty}^{\infty} d^2x \ e^{i {\vec q} \cdot {\vec x}} {\vec u}_{\perp}({\vec x})$. Here $\kappa (q)$ and $\mu (q)$, $\lambda(q)$ are the renormalized wave vector dependent bending and elastic moduli, respectively, and we assume the scaling forms
\begin{eqnarray}
\kappa (q) \sim q^{-\epsilon}, \ \mu (q) \sim \lambda (q) \sim q^{\omega}
\end{eqnarray}
with the exponents $\epsilon$ and $\omega$. It has been suggested that such singular behaviors in the effective moduli arise from the nonlinear coupling between phonon and undulation modes~\cite{Nelson1990}. It should be noted here that, for flexible self-avoiding polymerized membranes studied in the present paper, there is no explicit bending rigidity in the microscopic Hamiltonian (cf. the model of MC simulation). Nevertheless,  the effective bending rigidity is generated by excluded volume constraints between monomers by means of thermal fluctuations. Thus, $\kappa$ as well as $\mu$ and $\lambda$ should be proportional to temperature~\cite{Nelson1990}.

The amplitude of the undulation mode of the polymerized membrane with a linear size $L=Ma$, where $a (\sim b)$ is a short-wavelength cutoff, is calculated as
\begin{eqnarray}
\langle u_{\parallel}^2({\vec x})\rangle \simeq \frac{k_BT}{2\pi} \int_{\pi/L}^{\pi/a}\frac{q dq}{q^4 \kappa(q)} \simeq \frac{k_BT a^2}{\kappa_0}M^{2\zeta_{\parallel}}, 
\end{eqnarray}
where $\zeta_{\parallel}=1-\epsilon/2$ is a roughness exponent, whose numerical value is known to be $\zeta_{\parallel} \simeq 0.6$~\cite{Zhang, Petsche_Grest}. The bending modulus is denoted as $\kappa (q) = \kappa_0 (qa)^{-\epsilon}$ with $\kappa_0$ being the wave vector independent renormalized amplitude; below we adopt similar notations for the in-plane elastic moduli, too, i.e., $\mu (q) = \mu_0 (qa)^{\omega}$, $\lambda (q) = \lambda_0 (qa)^{\omega}$. The fact that $\zeta_{\parallel} < 1$ indicates the long range order of the surface normal. The flat phase is thus realized due to the wave-vector dependent effective bending modulus, which diverges for $q \rightarrow 0$.

The in-plane fluctuation can be calculated similarly as 
\begin{eqnarray}
\langle u_{\perp}^2({\vec x})\rangle \simeq (k_BT/(\lambda_0 + 2\mu_0)) M^{2 \zeta_{\perp}}
\end{eqnarray}
 with $\zeta_{\perp} = \omega/2$. The rotational invariance connects the in-plane and the out-of-plane scaling exponents through the relation\cite{Lubensky} 
\begin{eqnarray}
\omega = 2(1-\epsilon)  \Leftrightarrow \zeta_{\perp}=-1 + 2 \zeta_{\parallel} .
\label{parallel_perp_relation}
\end{eqnarray}

Next, let us define the terminal times $\tau_{\parallel}$ and $\tau_{\perp}$ for undulation and phonon modes, respectively, from the overall translational diffusion timescales $D \tau_{\parallel} \simeq \langle u_{\parallel}^2({\vec x})\rangle \sim L^{2 \zeta_{\parallel}}$ and $D \tau_{\perp} \simeq \langle {\vec u}_{\perp}^2({\vec x})\rangle \sim L^{2 \zeta_{\perp}}$. In the free-draining (Rouse) dynamics, the diffusion coefficients for each direction are given by $D_{\natural} \simeq k_BT/(\gamma_{\natural} a^2N)$, where $N \sim M^2$ is the total number of monomers and $\gamma_{\natural}$ is the friction coefficient in $\natural$ direction per unit area. (Here and in what follows, we use the subscript symbol $\natural$ to indicate either $\parallel$ or $\perp$ symbol to avoid repeating the same formula for each component.) This leads to the scaling laws
\begin{eqnarray}
\tau_{\natural} \simeq \tau_{0 \natural} \ M^{z_{\natural}}, 
\label{tau}
\end{eqnarray}
with the dynamical exponent 
\begin{eqnarray}
z_{\natural} = 2(1+\zeta_{\natural}),
\label{z_zeta}
\end{eqnarray}
 where $\tau_{0 \parallel} = \gamma_{\parallel} a^4/\kappa_0$ and $\tau_{0 \perp} = \gamma_{\perp} a^2/(\lambda_0 + 2 \mu_0 )$ are monomeric time scales in respective components.

\begin{figure}[h]
 \begin{center}
\includegraphics[width=0.4\textwidth]{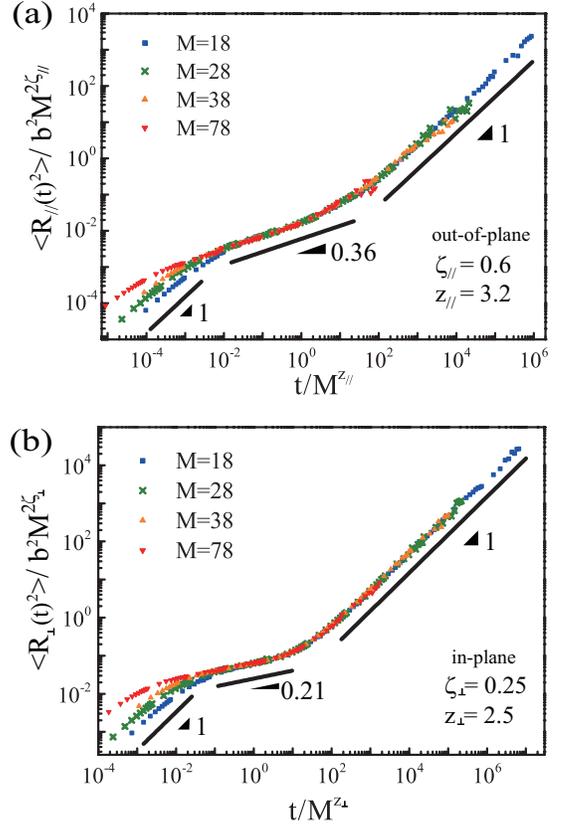}
\caption{Rescaled plots of Fig. 4. (a) the out-of-plane component, and (b) the in-plane component. In the data fitting, we adopt the scaling relation~(\ref{z_zeta}).}
\label{Fig5}
\end{center}
\end{figure}

Assuming the self-similar dynamical process, the typical displacement of each monomer in the $\natural$ direction may be written as 
\begin{eqnarray}
R_{\natural}(t) \simeq aM^{\zeta_{\natural}} \ \phi_{\natural} \left( \frac{t}{\tau_{\natural}}\right), 
\end{eqnarray}
where the scaling functions satisfy (i) $\phi_{\natural} (s)\simeq 1$ at $t \simeq \tau_{\natural}$ and (ii) $\phi_{\natural} (s) \simeq s^{\alpha_{\natural}}$ at $ \tau_{0 \natural} \ll t \ll \tau_{\natural}$. Requiring the motion to be independent of the system size $L$ in the intermediate time regime (cf. Fig.~\ref{Fig4}), the growth exponents can be deduced as $\alpha_{\natural} =\zeta_{\natural}/z_{\natural} $, or 
\begin{eqnarray}
\alpha_{\natural}=\frac{\zeta_{\natural}}{2(1+\zeta_{\natural})},
\label{alpha_exp}
\end{eqnarray}
using the relation~(\ref{z_zeta}).

We attempt to determine exponents $\zeta_{\natural}$ and $z_{\natural}$ by data collapsing the intermediate to large scale parts of Fig.~\ref{Fig4}. Given a limited range of system sizes, however, the error estimation in the two parameter fitting turns out to be difficult. We thus adopt the expected relation~(\ref{z_zeta}) and performed the data collapse, regarding $\zeta_{\natural}$ as a fitting parameter.  Figure~\ref{Fig5} shows excellent data collapses for the parameters $\zeta_{\parallel} = 0.6 \pm 0.05$ and $\zeta_{\perp} = 0.25 \pm 0.05$.

We thus find (i) the value of $\zeta_{\parallel}$ agrees with that found in previous static analysis measuring the roughness of the membrane without free edge effect~\cite{Zhang, Petsche_Grest} (see footnote~\ref{fn1}); (ii) the anomalous diffusion exponents $2 \alpha_{\parallel} \simeq 0.37$ and $2 \alpha_{\perp} \simeq 0.2$ predicted from the dynamical scaling relation~(\ref{alpha_exp}) is very close to those determined from the slope of the tagged monomer MSD in the out-of plane direction $2\alpha_{\parallel} = 0.36 \pm 0.01$ and that in the in-plane direction $2\alpha_{\perp} = 0.21 \pm 0.01 $; (iii) the relation~(\ref{parallel_perp_relation}) is confirmed to a good extent; (iv) the terminal time of the undulation mode is longer than that of the in-plane phonon mode, i.e., $z_{\parallel} > z_{\perp}$ as anticipated from Fig.~\ref{Fig3}; (v) large membranes maintain their orientation in the time range during which all the internal modes relax, i.e., $\tau_{\times} \gg \tau_{\parallel} \gg \tau_{\perp}$. This last point assures the validity of our protocol to measure the time course of the in-plane and the out-of-plane displacements of monomer in MC simulations using membrane orientation ${\vec n}(t_0)$ at time $t_0$.

\section{Generalized Langevin description}
Physically, the anomalous dynamics of the monomer is a manifestation of the collective effect of the internal modes present in the membrane.
To elucidate this point, we now proceed to the analysis of the internal modes, which leads to a generalized Lanvevin equation for the motion of a tagged monomer. A negative power-law memory kernel is shown to be generated through the superposition of the internal modes with a broad range of relaxation times~\cite{Panja_GLE, Sakaue}. 

The out-of-plane motion of monomers can be described by the following Langevin equation;
\begin{eqnarray}
\gamma_{\parallel} \frac{du_{\parallel}({\vec x}, t)}{dt}=&& - \frac{\delta {\mathcal H}\{u_{\parallel}({\vec x}, t), {\vec u}_{\perp}({\vec x})\}}{\delta u_{\parallel}({\vec x}, t)} \nonumber \\
&& \qquad \ +g_{\parallel}({\vec x}, t)+f_{\parallel}({\vec x}, t), 
\label{L_eq}
\end{eqnarray}
where $g_{\parallel}({\vec x}, t)$ is a white noise with zero mean and $\langle g_{\parallel}({\vec x}, t) g_{\parallel}({\vec x}', t')\rangle=2 \gamma_{\parallel} k_BT \delta^2({\vec x}-{\vec x}')\delta(t-s)$, and $f_{\parallel}({\vec x}, t)$ is a weak external force to probe the linear response. 
To analyze the motion of the tagged monomer, we manipulate a particular monomer labeled by ${\vec x}^*$ by external force $f_{\parallel}({\vec x}, t) = f_{0 \parallel} \delta^2({\vec x}- {\vec x}^*) U(t)$, where function $U(t)$ represents the time dependence of the protocol.
The coupled equations of motion (Eq.~(\ref{L_eq})) in real space can be decomposed into a set of normal modes in Fourier space
\begin{eqnarray}
 \gamma_{\parallel} \frac{d u_{\parallel}({\vec q}, t)}{dt} = -\kappa(q) q^4u_{\parallel}({\vec q}, t) + g_{\parallel}({\vec q}, t) + f_{\parallel}({\vec q}, t), 
 \label{L_eq_q}
\end{eqnarray}
where $\langle g_{\parallel }({\vec q}, t) \rangle =0$, $\langle g_{\parallel }({\vec q}, t) g_{\parallel }({\vec q'}, t') \rangle = 8\pi^2 \gamma_{\parallel} k_BT \delta^2 ({\vec q}+{\vec q}')\delta(t-t')$ and $f_{\parallel }({\vec q}, t) = f_{0 \parallel} e^{i {\vec q}\cdot {\vec x}^*} U(t)$.
Solving this equation, then, returning to the real coordinate, one can obtain the following generalized Langevin equation for the out-of-plane motion of the tagged monomer
\begin{eqnarray}
\frac{du_{\parallel}({\vec x}^*, t)}{dt} = \int_{-\infty}^t \mu_{\parallel}(t-s) f_{\parallel}(s) ds + \eta_{\parallel}(t), 
\label{GLE}
\end{eqnarray}
where $f_{\parallel}(s) = f_{0 {\parallel}} U(t)$ is the force in the out-of-plane direction acting on the tagged monomer, and the mobility kernel is $\mu_{\parallel}(t) = (2/\gamma_{\parallel}) \delta (t) + (2/\gamma_{\parallel} N) \delta (t) + \mu_{m \parallel}(t)$. The first term represents an instantaneous response of the monomer, which is responsible for the  short time scale ($t \lesssim \tau_{0 \parallel}$) normal behaviors, while the second term arises from the ${\vec q}={\vec 0}$ (center of mass) mode. The last non-local term originates from the coupling with the internal degrees of freedom, and can be explicitly written as
\begin{eqnarray}
\mu_{m \parallel}(t)  \simeq - \frac{1}{(2 \pi \gamma_{\parallel})^2 } \int d^2 q \ \kappa(q) q^4 \exp{\left(-\frac{\kappa(q) q^4}{\gamma_{\parallel}}t \right)} \nonumber \\
\simeq - \frac{ 1}{2 \pi  \gamma_{\parallel} a^2 \tau_{0 \parallel} } \int_{ \pi a/L}^{\pi} d p \ p^{3+2\zeta_{\parallel}} \ \exp{\left(-\frac{p^{2(1+\zeta_{\parallel})} \ t}{\tau_{0 \parallel}} \right)} .
\label{mu}
\end{eqnarray}
The argument in the exponential factor verifies the terminal time estimation in Eq.~(\ref{tau}), which corresponds to the relaxation time of the longest wavelength, i.e., $p = (2 \pi a/L)$.
The last term in Eq.~(\ref{GLE}) is interpreted to be a colored noise, for which one can verify the fluctuation-dissipation relation $\langle \eta_{\parallel}(t) \eta_{\parallel}(s)\rangle = k_BT\mu_{\parallel}(t-s)$~\cite{Panja_GLE, Sakaue}. 
For $\tau_{0 \parallel} \ll t \ll \tau_{\parallel}$,  the integral range in Eq.~(\ref{mu}) can be extended to infinity, yielding 
\begin{eqnarray}
\mu_{m \parallel}(t) &\simeq& -\frac{1}{4 \pi \gamma_{\parallel} a^2 \tau_{0 \parallel} (1+ \zeta_{\parallel})}  \nonumber \\
&\times&\Gamma \left(\frac{2+\zeta_{\parallel}}{1+\zeta_{\parallel}}\right) \left( \frac{t}{\tau_{0 \parallel}}\right)^{-(2+\zeta_{\parallel})/(1+\zeta_{\parallel})} , 
\label{mu_power}
\end{eqnarray}
where $\Gamma()$ is $\Gamma$-function. This negative power-law memory kernel  is a prominent feature in the connected systems, and dominates the motion of monomers at intermediate time scale ($\tau_{0 \parallel} < t < \tau_{\parallel}$), which is generally sub-diffusive.
From Eq.~(\ref{GLE}) and the fluctuation-dissipation relation, MSD of the tagged monomer in the out-of-plane direction can be calculated as 
\begin{eqnarray}
 \int_{t_0}^{t+t_0} dt_1\int_{t_0}^{t+t_0} dt_2 \left \langle\frac{du_{\parallel}({\vec x}^*, t_1)}{dt} \frac{du_{\parallel}({\vec x}^*, t_2)}{dt} \right \rangle \sim t^{2 \alpha_{\parallel}}
\end{eqnarray}
with the anomalous exponent given by Eq.~(\ref{alpha_exp})  in accordance with the preceding simple dynamical scaling argument. For the in-plane phonon mode, the analysis essentially parallels to that for the out-of-plane mode, leading to the memory kernel 
\begin{eqnarray}
\mu_{m \perp}(t) \sim - \left(\frac{t}{\tau_{0 \perp}}\right)^{-(2+\zeta_{\perp})/(1+\zeta_{\perp})}.
\end{eqnarray}
Thus, the anomalous exponent again given by Eq.~(\ref{alpha_exp}) for the in-plane motion.

\section{Hydrodynamic interaction}
While Eq.~(\ref{L_eq}) is the simplest form of the dissipative dynamics, where the mobility coefficient $\mu_{\natural}= \gamma_{\natural}^{-1}$ is a constant (free draining dynamics), the motion of an object in fluid generally induces the flow of solvents, which in turn affects the dynamics of the object. This so-called back flow effect couples the motions of distant monomers, and makes the mobility coefficient wave length dependent;
\begin{eqnarray}
\mu_{\natural}(q) = \gamma_{\natural}^{-1} + \mu_{\natural}^{(HD)}(q)
\label{mu_general}
\end{eqnarray}
Here, we shall briefly discuss how it modifies the dynamical exponents.
The form of the back flow contribution $\mu_{\natural}^{(HD)}$ depends on the dimension ($d=3$) of the embedding space, and the internal dimension (${\mathcal D}=2$) of the object as well as its conformation.
Based on the Kirkwood-Riseman approximation, one can estimate $\mu_{\natural}^{(HD)}$ for polymerized membranes in the flat phase as
\begin{eqnarray}
\mu_{\natural}^{(HD)}(q) \sim \frac{1}{\eta_0 q}, 
\label{mu_HD}
\end{eqnarray}
where $\eta_0$ is the viscosity of solvents~\cite{Frey}.
The scaling form $q^{-1}$ indicates that the hydrodynamic interaction is relevant in the large scale limit.
Using Eq.~(\ref{mu_HD}) in the equation of motion~(\ref{L_eq_q}), we find that the memory kernels become
\begin{eqnarray}
\mu_{m \natural} (t)  \sim - \left( \frac{t}{\tau_{0 \natural}}\right)^{-2(1+\zeta_{\natural})/(1+2\zeta_{\natural})}
\end{eqnarray}
The anomalous exponents for the tagged monomer diffusion thus become
\begin{eqnarray}
\alpha_{\natural} = \frac{\zeta_{\natural}}{1+2 \zeta_{\natural}}
\label{alpha_HD}
\end{eqnarray}

The same result can be obtained simply by noting that the diffusion coefficient of the flat object in fluid is inversely proportional to its linear dimension $D \sim M^{-1}$.
This contrasts to the case in the free draining dynamics $D \sim M^{-2}$. Following the same argument as before around  Eqs.~(\ref{tau}) to~(\ref{alpha_exp}), we find the dynamic exponents
\begin{eqnarray}
z_{\natural} = 1 + 2 \zeta_{\natural}
\end{eqnarray}
which again leads to Eq.~(\ref{alpha_HD}).

In real systems, the strength of the hydrodynamic interaction would depend on the solvent permeability of the membrane. Equation~(\ref{mu_general}) indicates that depending on the time and length scales, the crossover from the free-draining to the hydrodynamic interaction dominated dynamics may be expected as described in Ref.~\cite{Frey}. Note that the rotational diffusion time is also modified as $\tau_{\times} \sim M^3$.

\section{Summary}
Based on the results of MC simulations, we have constructed the dynamical scaling scenario for the motion of a tagged monomer in a polymerized membrane, distinguishing the two components of diffusion relative to the membrane; the out-of-plane and the in-plane diffusions. 
The diffusion exponents ($\alpha_{\parallel},  \ \alpha_{\perp}$) and the roughness exponents ($\zeta_{\parallel}, \ \zeta_{\perp}$) were estimated numerically, and their values were shown to be consistent with the dynamical scaling relations and with the relation imposed by the rotational symmetry between the bending and the elastic moduli. We also derived the generalized Langevin equations for a tagged monomer in a membrane to show that the system has negative memory kernels with a power law.
Finally, we discussed the effect of hydrodynamic interactions, focusing on how it alters the scaling exponents.
We believe that this is an important step toward understanding rheological properties of solutions containing polymerized membranes.

\acknowledgments
This work is supported by KAKENHI (No. 26103525,``Fluctuation \& Structure") and (No. 24340100) from  MEXT, Japan,
and JSPS Core-to-core Program, ``Non-equilibrium dynamics of soft matter and information''.

\end{document}